\documentclass[sigconf]{acmart}

\renewcommand\footnotetextcopyrightpermission[1]{}
\usepackage{threeparttable}
\usepackage{listings}
\usepackage{xcolor}
\usepackage{float}
\usepackage{makecell}
\usepackage{enumitem}
\usepackage{bbding}   

\begin{document}

\title{Coati: Host-Grounded Execution for Installable Mobile GUI Agents}


\renewcommand{\shortauthors}{xx et al.}


\author{Yudou Ying}
\authornotemark[1]
\affiliation{%
  \institution{Hangzhou Institute for Advanced Study, University of Chinese Academy of Sciences}
  \city{Hangzhou}
  \country{China}
}

\author{Pu Shi}
\authornote{Both authors contributed equally to this research.}
\affiliation{%
  \institution{Independent Researcher}
  \country{China}
}

\author{Hanlei Shi}
\affiliation{%
  \institution{Hangzhou Institute for Advanced Study, University of Chinese Academy of Sciences}
  \city{Hangzhou}
  \country{China}
}

\author{Yuhua Zheng}
\authornote{Corresponding author.}
\affiliation{%
  \institution{Hangzhou Institute for Advanced Study, University of Chinese Academy of Sciences}
  \city{Hangzhou}
  \country{China}
}

\renewcommand{\shortauthors}{Shi et al.}

\begin{abstract}
Mobile GUI agents can carry out tasks across apps, yet most run on a separate
computer that drives a phone over the Android Debug Bridge (ADB), out of reach of ordinary users
and of the apps through which most people reach Web services. We present \textbf{Coati}, a
source-available agent that users install as an Android app. Its runtime on the phone, the host, records
what a model alone cannot see: the effect of each action at its target, result screens across
apps, and the progress of repeated steps. \textbf{VerifyLoop} checks action outcomes and completion claims against host evidence and reports
a completion it cannot audit as unverified. \textbf{AppMemory} builds per-app procedural memory
from observations that the interface supports and reuses it on later visits. A \textbf{routed fast
executor} lets a lightweight selection model take simple, confident steps and lets the host repeat
loops of simple steps that the large model plans once. On all 116 AndroidWorld tasks, with Gemini
3.8 Flash as every agent's backbone, Coati succeeds on 86.5\% of the tasks over three trials, ahead of
the strongest open-source agent Artemis (77.6\% with the same accessibility-only access, 76.5\% even
when allowed ADB) and far ahead of FinalRun and Midscene (about 59\%), and VerifyLoop cuts false
completion claims by about a third at comparable success. In everyday use, the fast executor cuts
the median time of phone tasks by 39\% at comparable success and runs three of four loop tasks
2.8--5.4$\times$ faster with 6--9$\times$ fewer model tokens, while AppMemory shortens a repeated
deep route in an unfamiliar app from 18.5 steps to 6--7. On an Android~16 phone, routed execution
completes all 27 benchmark runs with 76\% fewer Reasoning calls on the critical path. Host evidence, not model self-assessment,
makes an installable agent reliable and fast.
\end{abstract}

\ccsdesc[500]{Computing methodologies~Artificial intelligence}
\ccsdesc[500]{Computing methodologies~Planning and scheduling}
\ccsdesc[500]{Computing methodologies~Knowledge representation and reasoning}
\ccsdesc[300]{Human-centered computing~Ubiquitous and mobile computing systems and tools}
\keywords{Mobile GUI Agents, Multimodal Large Language Models, Model Routing, Screen Parsing,
On-device OCR, Execution Verification, Agent Memory}

\maketitle

\section{Introduction}
\label{sec:intro}
Multimodal large language models can now operate graphical user interfaces (GUIs) directly: they read
a screenshot or an interface tree and act through taps, swipes and text
entry~\cite{hong2024cogagent,zheng2024seeact,nguyen2025guiagents}. On mobile devices this matters
more than anywhere else, because most applications expose no programmable interface and their
layouts change across versions, languages and device
states~\cite{rawles2023androidwild,rawles2025androidworld}. People also access many Web services
through mobile applications. A mobile GUI agent intended for everyday use should therefore run directly
on the phone and operate these applications.

However, most mobile GUI agents run on a separate computer that drives the phone over the Android
Debug Bridge (ADB)~\cite{appium,rawles2025androidworld}. The computer sends actions to the phone and receives observations in return. For ordinary users, installing and operating such a system can be difficult. More importantly, the agent can understand the phone only through the information returned by the bridge. Therefore, it may miss fine-grained changes after an action, making it difficult to determine whether the action succeeded and whether the task actually reached its target state~\cite{lee2025verisafe,wu2025backtrackagent,zhang2026verigui}.
These limitations suggest a more direct and promising deployment model: an agent installed directly on the phone. Because its runtime and the interface it operates reside on the same device, the agent can observe the interface state and action outcomes in real time. It can also maintain records of cross-application results and the progress of repeated actions.

Running an agent directly on the phone also imposes practical requirements. Each step should avoid unnecessary waiting and model calls. Since users may not continuously monitor the screen, a task reported as complete must actually reach the requested state. Repeated tasks should not require the agent to rediscover the same route each time. Existing mobile GUI agents do not fully satisfy these requirements. They typically ask a large model to make decisions step by step, rely on the model to judge whether a task is complete, and treat repeated actions as separate model decisions. These choices can increase latency and token costs, while making execution failures more difficult to detect~\cite{wang2024mobileagentv2,wen2024autodroid,li2025mobileuse}.
A mobile agent must therefore use local evidence not only to execute actions, but also to verify their
outcomes, assess task completion, and retain useful experience. This calls for a different execution
principle: the model may propose an action, but the runtime on the phone should determine what
actually happened from evidence recorded by the host~\cite{lee2025verisafe,wu2025backtrackagent,zhang2026verigui}.

We present \textbf{Coati}, a source-available Android agent that users install as an app. Coati is designed to address the reliability, execution-efficiency, and repeated-use challenges of installable mobile agents. Its core principle is \emph{host-grounded execution}: the model proposes an action, while evidence recorded by the host---the information collected by Coati's runtime on the phone---determines the outcome.

We implement this principle with three mechanisms. \textbf{(1)VerifyLoop} checks each action against the
interface changes it causes and audits completion claims using the result screens of the applications
involved and the progress of delegated loops. When the available evidence is insufficient, it reports
the task as unverified rather than complete. \textbf{(2)AppMemory} stores interface observations as
per-application guidance only when they are supported by the interface and the task passes the
completion audit. It then retrieves this guidance during later visits. 
\textbf{(3)Routed fast executor} reduces the cost of model calls. A lightweight typed selector handles
simple, confident actions from host-built candidates, while the Reasoning model handles uncertain
steps, application switches, and text entry under a user-selectable speed--accuracy mode. For
repeated simple actions, the Reasoning model plans the loop once, and the host executes it item by
item while recording the handled items. Together, these mechanisms use the same host evidence to
improve execution efficiency and constrain what the agent can claim and remember.

We compare Coati with the strongest open-source agents on all 116 AndroidWorld tasks. All agents
use Gemini~3.8~Flash as the backbone model and run on the same device with the same task instances
and execution protocol. Coati achieves an average task success rate of 86.5\% across three trials,
ahead of the strongest open-source agent Artemis (77.6\% with the same accessibility-only access,
and 76.5\% even when allowed ADB), and far ahead of FinalRun and Midscene (about 59\%). At
comparable success rates, VerifyLoop reduces false completion claims by about one third
(Section~\ref{subsec:rq1}--\ref{subsec:rq2}).
Beyond AndroidWorld, whose tasks are short first visits, we evaluate execution efficiency and
repeated use in everyday settings. Compared with a per-step Reasoning configuration, the routed
fast executor reduces the median execution time of phone tasks by 39\% at comparable success. It
also makes three of four loop tasks 2.8--5.4$\times$ faster while reducing model token usage by
6--9$\times$. For repeated deep navigation in an unfamiliar application, AppMemory reduces the
average execution length from 18.5 steps to 6--7 steps
(Section~\ref{subsec:rq3}--\ref{subsec:rq4}).

Overall, Coati unifies action verification, execution acceleration, and experience reuse within a
single host-evidence loop. Our evaluation examines these mechanisms from four perspectives:
benchmark capability, completion reliability, everyday execution speed, and repeated use. The results
show that improving the reliability and efficiency of an installable mobile agent depends not only on
the model itself, but also on whether the host can continuously record and use evidence from actual
execution. Our contributions are:
\begin{itemize}[leftmargin=*, labelsep=0.5em]
\item \textbf{A host-grounded execution framework for mobile GUI agents.} The framework uses host
evidence, rather than model self-assessment, to decide if an action succeeded, if a task was completed, and what is worth remembering.

\item \textbf{Three core mechanisms.} We design and implement VerifyLoop, AppMemory, and a routed fast
executor with loop delegation, and integrate them into an installable Android application. The source
code is available for noncommercial use.

\item \textbf{An evaluation that distinguishes benchmark capability from everyday use.} We conduct
three trials on all 116 AndroidWorld tasks with the same backbone model, comparing against the
strongest open-source agent Artemis as well as FinalRun and Midscene, and use everyday tasks, simple
action loops, and repeated deep-navigation tasks to evaluate the practical contribution of each
mechanism.
\end{itemize}

\section{Related Work}
\subsection{Mobile GUI Agents and Their Deployment}
Mobile GUI automation began with scripts and demonstrations, such as Appium's
selectors~\cite{appium}. Multimodal models turned it into general-purpose agents:
SeeClick~\cite{cheng-etal-2024-seeclick}, Aria-UI~\cite{yang-etal-2025-aria} and
UI-TARS~\cite{qin2025uitars} improve grounding, while AndroidGen~\cite{lai-etal-2025-androidgen},
Mobile-Agent~\cite{wang2024mobileagent,wang2024mobileagentv2} and
Mobile-Agent-v3~\cite{ye2025mobileagentv3} combine screenshots, interface information and history for
multi-step tasks; pure-vision parsers such as OmniParser~\cite{lu2024omniparser} and accessibility-based
agents~\cite{dai2026vdroid} cover complementary screens. AndroidWorld~\cite{rawles2025androidworld}
provides the standard dynamic benchmark. Training-based work scales environments, data and
reinforcement learning: MobileRL~\cite{xu2026mobilerl} trains with online agentic reinforcement
learning, and OpenMobile-2~\cite{yan2026openmobile2} trains on more than 110 apps with generated
trajectories and app-native tools. Such work raises a single model's capability but still calls that
model at every step and does not accumulate application knowledge in use; Coati's routing and
AppMemory act at run time and are complementary to it. Nearly all of these agents drive the phone
from a separate computer and ask a large model for every step. Coati runs on the phone, and the
evidence it records there is what its decisions rest on.

\subsection{Verification, Reflection and Experience}
Mobile-Agent-v2~\cite{wang2024mobileagentv2} and MobileUse~\cite{li2025mobileuse} use reflection to
detect and recover from erroneous actions; VeriSafe Agent~\cite{lee2025verisafe} and
V-Droid~\cite{dai2026vdroid} verify candidate actions before execution, and
BacktrackAgent~\cite{wu2025backtrackagent} and VeriGUI~\cite{zhang2026verigui} study error
recovery. These verifiers ask a model to judge the model's own progress. VerifyLoop decides with host
evidence wherever it exists (pixel and text changes at the target, result screens across apps, loop
progress) and uses a model only to read that evidence. For experience,
AutoDroid~\cite{wen2024autodroid}, AppAgent~\cite{zhang2025appagent,li2024appagentv2},
GUI-explorer~\cite{xie2025guiexplorer}, LLM-Explorer~\cite{zhao2025llmexplorer} and
AppCards~\cite{li2026curiosity} build application knowledge from exploration, documentation or
trajectories; AppMemory determines which observations from real tasks are reusable and retains only those supported by the interface.

\subsection{Fast and Slow Decision Making}
Dual-process designs pair a fast policy with a slower deliberate one~\cite{lin2023swiftsage}, and
model cascades and routers send easy queries to cheaper models~\cite{chen2023frugalgpt,ong2024routellm}.
These approaches typically use query difficulty, model confidence, or expected cost to decide when
additional reasoning is necessary. In GUI agents, however, the difficulty of a step also depends on
the current interface state: a simple tap, a text-entry action, an application switch, and an
irreversible commit may require different levels of deliberation even within the same task. Existing
routing methods generally select a model for a request, while leaving target resolution, action
execution, and post-action verification to the same reasoning pipeline.
Coati routes at the level of GUI steps: the fast route selects among candidates the host can execute
exactly instead of generating actions, routing follows host-observable state (verification outcomes,
exploration, pending commits), and repetitive work is delegated as a loop that the host executes and
tracks, rather than as a sequence of model decisions.

\section{Host-Grounded Execution}
\label{sec:methods}

\subsection{Execution Setting and Formalization}
\label{subsec:problem-definition}

We consider a GUI agent operating on a mobile device. Given a natural-language task goal $g$, the
agent observes the current application state and completes the task using general-purpose GUI
actions, such as tapping, swiping, text input, system-key operations, and application launching.
At execution step $t$, the observation is
\begin{equation}
    o_t=(I_t,U_t,p_t,f_t),
\end{equation}
where $I_t$ is the screenshot, $U_t$ the accessibility UI tree (possibly empty or partial), $p_t$ the
foreground application, and $f_t$ the input-focus state. A perception function turns $o_t$ into a
parsed screen $S_t=P(o_t)$: a set of elements with text, type, bounds, supported operations and state.

For application $i$, the agent may retrieve task- and interface-relevant experience
$m_t=R(M_i,g,U_t)$ from its Living Manual $M_i$. Coati decides each step on one of two routes. A router $\rho$ selects the route from the execution
state $x_t$ (verification outcomes, exploration state, candidate confidence, pending commits):
\begin{equation}
    \rho(x_t)\in\{\textsc{Fast},\textsc{Reason}\}.
\end{equation}
On the \textsc{Fast} route, the host builds a candidate set $\mathcal{A}_t=\Gamma(S_t)$ and a typed
selector $\sigma$ chooses one candidate together with a verdict $\hat v_{t-1}$ on the previous action:
\begin{equation}
    (a_t,\hat v_{t-1})=\sigma(g,S_t,\mathcal{A}_t,h_t,m_t,\kappa_t),
\end{equation}
where $\kappa_t$ is background guidance (sub-goals, hints, text values) produced asynchronously by
the Reasoning model. On the \textsc{Reason} route, the Reasoning model produces a structured action,
its expected outcome $e_t$, a completion claim $d_t$, and an optional page observation $z_t$:
\begin{equation}
    (a_t,e_t,d_t,z_t)=\pi(g,o_t,h_t,m_t).
\end{equation}
On both routes Coati makes two further decisions. It evaluates whether the transition caused by $a_t$
matches what was expected,
\begin{equation}
    v_t=V(o_t,a_t,e_t,o_{t+1}),
\end{equation}
and whether a page observation $z_t$ is supported and reusable,
\begin{equation}
    r_t=A(z_t,o_t,id_t,M_i),
\end{equation}
where $id_t$ is the resolved identity of the foreground application. $V$ evaluates an
action-induced state transition, whereas $A$ evaluates the evidential support and reusability of
facts observed on the current page; $v_t$ therefore does not determine $r_t$. Admitted observations
accumulate in a candidate set $\mathcal{C}_i$ and update the manual only after the task is accepted
as completed, $M_i'=U(M_i,\mathcal{C}_i)$.

Throughout, \emph{host evidence} $\epsilon_t$ denotes what the agent records itself rather than
receives from a model: the accessibility tree at the time of the action, pixel and text changes at
and around the target, the last result screen of every application the task used, and the record of
a delegated loop (Section~\ref{subsec:routing}). $V$, the completion audit and $A$ are all evaluated
against $\epsilon_t$.

\begin{figure*}[t]
    \centering
    \includegraphics[width=\linewidth]{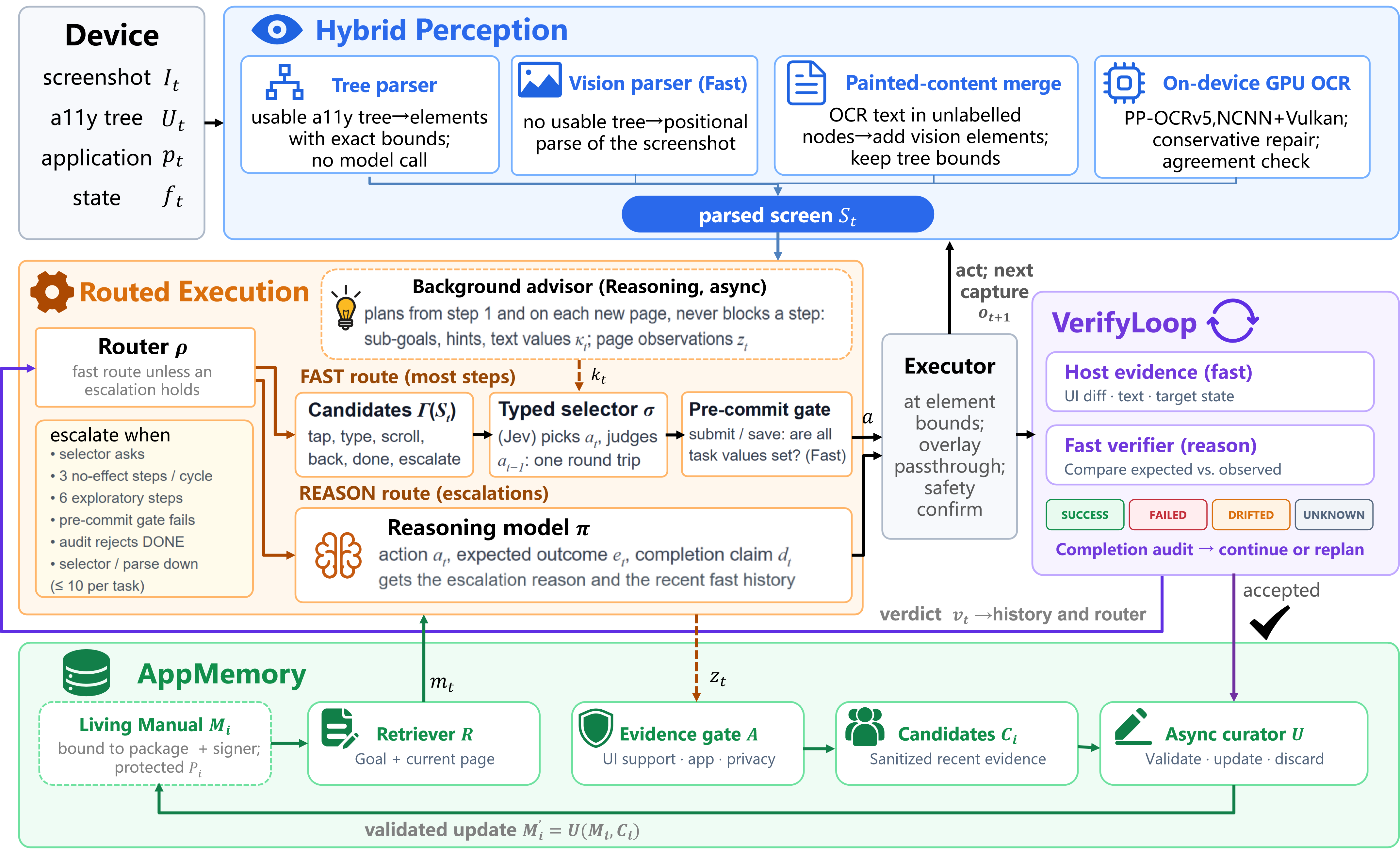}
    \caption{Overview of Coati. Hybrid perception produces a parsed screen from the accessibility tree, vision, or both. The router selects fast or Reasoning execution. VerifyLoop checks action outcomes using host evidence, while AppMemory provides advisory entries and updates the manual after accepted tasks.}
    \label{fig1}
\end{figure*}

\subsection{Coati Overview}
\label{subsec:coati-overview}

Figure~\ref{fig1} presents the architecture. At each step, Coati captures the screen and accessibility tree, and hybrid perception produces a parsed screen. The router selects either the fast selector or the Reasoning route. VerifyLoop checks each action and feeds its verdict to the next decision, while AppMemory retrieves advisory entries and learns only from accepted tasks. Both routes share capture, safety handling, completion auditing and task state.

Coati is an Android application whose accessibility service reads the interface and performs actions. An optional overlay shows progress and lets the user pause or stop the task. Permission and installation dialogs remain under user control.

\subsection{Hybrid Perception}
\label{subsec:perception}

Coati first uses the accessibility tree when it contains enough labelled, visible and actionable nodes. Each node becomes an element with text, type, bounds, supported operations and state, while containers inherit useful labels from their descendants. For custom canvases, games and WebViews without semantics, a Fast model parses the screenshot into a compact positional array of regions and elements with normalized coordinates, text, function, operations and state; this keeps a parse around one to two thousand output tokens. On-device GPU OCR is restricted to unlabelled, non-image nodes to recover painted widgets embedded in otherwise native screens. OCR may add or snap text and the system rejects a parse that describes another image, but it never overwrites model text. Detailed parse-level measurements appear in Appendix~\ref{app:perception}.

\subsection{Routed Execution}
\label{subsec:routing}

\textbf{\textit{Candidates and background planning.}} The host builds a finite candidate set from the parsed screen: operations on actionable elements, scrolling, back, app return or launch, waiting, completion and escalation. Each candidate carries the element text, type, state and exploration annotations, and the host executes it at the element bounds, so the fast route never generates coordinates. The selector receives the goal, background sub-goals and hints, retrieved manual entries, recent history, candidates and evidence about the previous action. One request chooses the next candidate, judges the previous effect and, when needed, supplies a typing value, so a healthy step costs one round trip. 
The Reasoning model plans at task start and on each new page, returning sub-goals, hints, required text and optional AppMemory observations. The fast route waits for this guidance only when it lacks a required text value.

\textbf{\textit{Escalation.}} The router uses the Reasoning route when the selector requests it, fast actions repeatedly fail or exploration remains uncertain, a commit is reached before the task state is complete, the completion audit rejects a claim, perception or selection is unavailable, or the step leaves the current app or enters text. It passes the reason and recent fast history to the Reasoning model, then resumes the fast route unless repeated escalations make reasoning the default for the remainder of the task.

\textbf{\textit{Speed--accuracy modes.}}
Whether a fast choice is trusted depends on its confidence and on whether the selector is exploring
or goal-directed. Coati exposes the thresholds as a user setting with three modes: \textsc{Fast}
executes every goal-directed choice and explorations above 0.4 on the fast route and also types text
there; \textsc{Balanced}, the default, raises the exploration threshold to 0.6 and sends text entry
to the Reasoning route; \textsc{Accurate} runs only goal-directed choices with confidence of at
least 0.8 on the fast route.

\textbf{\textit{Loop delegation.}} For repeated actions over list items, the Reasoning model plans the per-item operation, optional count, stop condition and scope once. The host then resolves the control on the first item, applies it to subsequent matching items without a model call per item, and records an item only when verification shows a target change. It scrolls when no unhandled match is visible and stops at the count, condition, list end or a bounded no-progress limit. The loop record—action, handled items, stop reason and list coverage—becomes host evidence, so a task over every matching item is not accepted when the loop did not reach the list end.

\textbf{\textit{Pre-commit gate.}}
Because the selector treats background guidance as a hint, it may commit a form whose values are not
all set, for example submitting an order with two copies when the task asks for three. Undoing a
commit costs several steps and an audit rejection. Before tapping a button, item or card whose label
denotes a commit (submit, save, send, export, confirm, pay, and their equivalents), Coati asks the
Fast model once whether the screenshot shows every value the task requires; if not, the step is
escalated with the missing value named. Tabs and switches are excluded, and a failed check never
blocks the action. Commits are rare, so the gate costs about one Fast call per task.

\subsection{VerifyLoop: Verification-Driven Mobile Execution}
\label{subsec:verifyloop}

Traditional Observe--Think--Act agents proceed to the next reasoning step immediately after issuing
an action. A successful return from the Android interface only indicates that the operation was
accepted; undetected failures, unexpected pop-ups, and page drift may propagate, and repeated
attempts under an unchanged interface may lead to unproductive loops. VerifyLoop extends execution to
\textbf{Observe--Think--Act--Verify} and converts verification outcomes into control signals.

On the Reasoning route, after target resolution and local safety checks the runtime executes $a_t$
through Android's AccessibilityService, $o_{t+1}=T(o_t,a_t)$, and a lightweight Fast model compares
the pre-action observation, the post-action observation and the expected outcome:

\begin{equation}
\begin{split}
    v_t &= V(o_t,a_t,e_t,o_{t+1}) \in \{\texttt{SUCCESS},\texttt{FAILED},\\
    &\quad \texttt{DRIFTED},\texttt{UNEXPECTED},\texttt{UNKNOWN}\}.
\end{split}
\end{equation}

\begin{table}[t]
\centering
\caption{Verification outcomes and corresponding control decisions in VerifyLoop.}
\label{tab:verify-outcomes}
\begin{tabular}{p{0.20\linewidth}p{0.32\linewidth}p{0.34\linewidth}}
\toprule
Outcome & Interpretation & Control Decision \\
\midrule
\texttt{SUCCESS} & The expected transition is observed. & Continue from the new state. \\
\texttt{FAILED} & The action is ineffective or the expected state is absent. & Return failure feedback and replan. \\
\texttt{DRIFTED} & Execution enters a task-incompatible state. & Recover from the observed state or terminate. \\
\texttt{UNEXPECTED} & The interface changes differently from the prediction. & Continue reasoning from the observed state. \\
\texttt{UNKNOWN} & Available evidence is insufficient. & Re-observe without assuming success. \\
\bottomrule
\end{tabular}
\end{table}

As summarized in Table~\ref{tab:verify-outcomes}, VerifyLoop maps each outcome to a distinct control
response. The result is incorporated into the execution history $h_{t+1}=H(h_t,a_t,e_t,o_{t+1},v_t)$,
which conditions the next step.

\textbf{\textit{Host-evidence verification on the fast route.}}
On the fast route, verification uses no extra model call. The host compares the two frames on a
luminance grid decoded at half resolution, locates changed cells relative to the acted element,
compares element texts, and reads the target's state (for example unselected to selected, a switch
turned on, typed text now visible). It phrases an expectation from the action type and asks the
selector to judge the previous action in the next step's request. A certain host verdict overrides the
selector: a transition is \texttt{NO\_CHANGE} only if no grid cell changed at all, while any change on
or beside the target, however small, counts as evidence (a switch knob moving, a counter next to a
``+'' button).

\textbf{\textit{Failure-aware stopping.}}
An action that keeps failing on the same target must stop, but a loose criterion ends tasks that
are still progressing. Let $\phi(a_t)$ be an action's type and target (its element, or its coordinate
cell at 4\% of the screen) and $\sigma(o_t)$ the screen signature. VerifyLoop counts the
\emph{consecutive} failed attempts of the current action,

\begin{equation}
\begin{split}
N_t &= \max\bigl\{n : \phi(a_{t-k})=\phi(a_t)\land\sigma(o_{t-k})=\sigma(o_t)\\
&\quad \land v_{t-k}\in\{\texttt{FAILED},\texttt{DRIFTED}\},\ 0\le k<n\bigr\},
\end{split}
\end{equation}

Waits and screenshots neither count toward nor reset the run, while any successful action resets it.

\textbf{\textit{Completion assessment.}} When a route claims completion, VerifyLoop runs a bounded audit $c_t=C(g,o_t,\epsilon_t)$ with a Fast model. It compares the current screen, the last result screen from every other application used, and any delegated-loop record against the values specified by the task. A rejected claim returns to execution and sends the reason to the Reasoning route; an unavailable audit ends the task with an explicit unverified status rather than plain success. For information-seeking tasks, the audit also reads the answer from the final screen.

\subsection{AppMemory: Evidence-Bounded App Experience}
\label{subsec:appmemory}

AppMemory maintains a \emph{Per-App Living Manual} of reusable page landmarks, functional relations and recovery instructions tied to the application identity and interface encountered on the user's device. This avoids treating generic documentation or complete trajectories as authoritative when versions, enabled features, dynamic content or sensitive values differ.

\textbf{\textit{Manual Initialization and Retrieval.}}
For each application $i$, AppMemory maintains a Living Manual $M_i$ bound to the application identity
$id_i=(pkg_i,sig_i)$, with the observed version and locale stored as compatibility metadata:
\begin{equation}
M_i^{(0)}=
\begin{cases}
S_i, & \text{if a package-bound seed exists},\\
\operatorname{Empty}(id_i), & \text{otherwise},
\end{cases}
\end{equation}
where a seed may define protected sections $P_i$ that subsequent online updates cannot overwrite.
Before each step, $m_t=R(M_i,g,U_t)$ retrieves an application overview and a bounded set of page
entries relevant to the task and the current screen; the overall retrieval and update process is
illustrated in Fig.~\ref{fig:appmemory} in Appendix~\ref{app:phone}.
Retrieved experience is advisory: the current screen, VerifyLoop's result and the safety policy always take precedence. On the fast route the same
entries are given to the selector as an app guide.

\textbf{\textit{Evidence-Bounded Admission.}}
In learning mode, page observations $z_t$ are evaluated by
\begin{equation}
r_t = A(z_t,o_t,id_t,M_i),\qquad
E_t = E_{\mathrm{ui}}\land E_{\mathrm{id}}\land E_{\mathrm{stable}}\land E_{\mathrm{privacy}},
\end{equation}
requiring visible interface support, correct application ownership, cross-task stability, and privacy
compliance (Table~\ref{tab:appmemory-decisions}). Only an observation classified as
\texttt{NEW\_OR\_CHANGED} and satisfying all conditions is admitted:
$\mathcal{C}_i\leftarrow\mathcal{C}_i\cup\{\operatorname{Sanitize}(z_t)\}$ if
$r_t=\texttt{NEW\_OR\_CHANGED}\land E_t$.

\begin{table}[t]
    \centering
    \small
    \caption{AppMemory admission and update decisions.}
    \label{tab:appmemory-decisions}
    \begin{tabular}{@{}p{0.19\columnwidth}p{0.29\columnwidth}p{0.40\columnwidth}@{}}
        \toprule
        \textbf{Decision} & \textbf{Condition} & \textbf{Action} \\
        \midrule
        \texttt{MATCHED} & Existing fact & Keep $M_i$ unchanged. \\
        \texttt{NEW} & New/changed fact and $E_t=1$ & Sanitize and add to $\mathcal{C}_i$. \\
        \texttt{UNKNOWN} & Insufficient evidence & Discard the observation. \\
        \midrule
        \texttt{NO\_CHANGE} & No valid difference & Preserve $M_i$. \\
        \texttt{EDIT} & Bounded additions & Apply a constrained local edit. \\
        \texttt{SUMMARIZE} & Redundancy or capacity limit & Reorganize under validation. \\
        \bottomrule
    \end{tabular}
\end{table}

\textbf{\textit{Controlled Manual Evolution.}}
Candidates remain task-local. After the task is accepted as completed, AppMemory asynchronously
builds a candidate manual and commits it only if validation succeeds:
\begin{equation}
\widehat{M}_i = U(M_i,\mathcal{C}_i),\qquad
M_i' =
\begin{cases}
\widehat{M}_i, & \operatorname{Valid}(\widehat{M}_i,M_i,P_i)=1,\\
M_i, & \text{otherwise}.
\end{cases}
\end{equation}
Validation checks structure, consistency, capacity, revision conflicts and protected sections; cancelled or invalid updates preserve the previous manual. Admission remains separate from VerifyLoop: an expected transition does not make every page fact reusable, and a failed action does not make those facts wrong.

\section{Evaluation}
\label{sec:experiments}

We ask four questions, one per claim:
\textbf{RQ1} (capability): with the same backbone model, how does Coati compare with open-source
agents on AndroidWorld?
\textbf{RQ2} (reliability): does host-grounded verification keep Coati from claiming success it has
not achieved?
\textbf{RQ3} (speed in everyday use): does the routed fast executor make everyday tasks and loops of
simple steps faster and cheaper without losing success?
\textbf{RQ4} (repeated use): does AppMemory shorten tasks that users repeat?

\subsection{Setup}
\label{subsec:experimental-settings}

\textbf{\textit{AndroidWorld.}}
AndroidWorld~\cite{rawles2025androidworld} has 116 parameterized tasks across 20 applications (61
easy, 36 medium, 19 hard), each with an official step budget and an evaluator that inspects the
resulting app state rather than the agent's claim. Every agent runs in the same Android~13 emulator
(Pixel~6 profile) with Gemini~3.8~Flash through Google's API as its backbone, on the same task
instances (seed 30), budgets and official evaluators. A harness starts each task, waits for the agent,
submits the answer of information tasks and calls the evaluator; the emulator is cold-restarted
every ten episodes. Coati runs in its per-step configuration (C2: the Reasoning route on every step,
VerifyLoop on) for three trials; the ablation C1 turns VerifyLoop off. The open-source agents
Artemis~\cite{artemis2026}, FinalRun~\cite{finalrun2026} and Midscene.js~\cite{midscene2026} run with
their released code through the same harness, with their model calls redirected to the same backbone;
Artemis is run both in its accessibility-only (GUI) mode, matching Coati's access, and, for one trial,
with the ADB access its release expects, while Midscene's default thinking level is rejected by the
API, so we use the lowest accepted level. Episodes in which the environment
failed before the agent acted are excluded and rerun under the same rule for every agent
(Appendix~\ref{app:protocol}). We report the success rate (SR, partial scores as given by the
evaluators), pass$^3$ (tasks solved in all three trials), the false-completion rate (declared
completions that the evaluator judged failed), and time and model tokens per episode.

These agents report far higher success in their own evaluations (around 90\% on this benchmark), and
two deliberate choices account for the gap here, neither of them the backbone model. First, every
agent is held to AndroidWorld's official per-task step budget (about ten steps for a simple task),
whereas the opponents' released configurations plan up to roughly 120 steps; under the official budget
most of their failures here are step-limit cut-offs rather than wrong decisions. Second, several of
their reported successes drive apps through ADB and root SQL writes to private databases; we compare
every agent through the on-screen interface only and do not credit ADB-won tasks, since an installed
consumer agent has no such access. The gap is not a stronger model on their side: the leaderboard's
near-perfect entries include a Gemini~3.7 submission, a Flash-tier backbone no stronger than the
Gemini~3.8~Flash every agent uses here. We therefore read Table~\ref{tab:androidworld-controlled} as a
controlled, standardized comparison under one step budget and one interface, not as a reproduction of
each agent's headline figure.

\textbf{\textit{Everyday use.}}
RQ3 and RQ4 use an emulator of the same profile with Google's stock applications and two open
fixture apps whose state can be checked programmatically (Appendix~\ref{app:tasks}). \emph{Classic}
is Coati with every step on the Reasoning route; \emph{Routed} adds the fast executor (selector
Jev~\cite{typesafejev}, version jev-1.13.0, \textsc{Balanced} mode). The two configurations alternate
within every round, so that server load affects both equally. Time is the agent's own duration from
task start to completion; tokens include background calls.

\subsection{Capability on AndroidWorld (RQ1)}
\label{subsec:rq1}

\begin{table}[t]
    \centering
    \small
    \caption{AndroidWorld, all 116 tasks, Gemini~3.8~Flash as every agent's backbone, same emulator,
    task instances, budgets and evaluators. SR: success rate (\%), mean over trials. False done:
    share of declared completions that the evaluator judged failed (\%). Time: mean seconds per episode;
    tokens: mean thousands per episode, as reported by each agent's model API. Midscene, FinalRun
    and Artemis~(ADB) drive the phone over ADB; Coati and Artemis~(GUI) use the accessibility API
    only. A dash is a value the agent does not report through the harness.}
    \label{tab:androidworld-controlled}
    \setlength{\tabcolsep}{3.2pt}
    \begin{tabular}{@{}lcccccc@{}}
        \toprule
        \textbf{Agent} & \textbf{Trials} & \textbf{SR} & \textbf{pass$^3$} & \textbf{False done} & \textbf{Time} & \textbf{Tokens} \\
        \midrule
        Midscene.js & 1 & 59.1 & -- & 7.5 & 179 & 222 \\
        FinalRun & 1 & 59.9 & -- & 14.5 & 224 & 177 \\
        Artemis~(ADB) & 1 & 76.5 & -- & -- & 185 & -- \\
        Artemis~(GUI) & 3 & 77.6 & 69.8 & -- & 209 & -- \\
        \textbf{Coati} & 3 & \textbf{86.5} & \textbf{81.9} & \textbf{5.4} & 172 & 154 \\
        \bottomrule
    \end{tabular}
\end{table}

Table~\ref{tab:androidworld-controlled} compares the agents under identical conditions. Coati
reaches 86.5\% on average (89.7, 87.1 and 82.8 across the three trials, sample standard deviation
3.5) and solves 95 of the 116 tasks (81.9\%) in all three trials. The strongest opponent, Artemis,
reaches 77.6\% over three trials with the same accessibility-only access, and 76.5\% in one trial
even when allowed ADB; FinalRun and Midscene solve 69.5 (59.9\%) and 68.5 (59.1\%) in their trial.
Coati thus leads the fairest comparison, Artemis without ADB, by 8.9 percentage points, leads
Artemis-with-ADB by ten while using no ADB itself, and leads FinalRun and Midscene by about 27, with
fewer tokens per episode than any and a mean time among theirs. The lead is not from claiming success
more readily (5.4\% false completions, against 14.5\% and 7.5\% for FinalRun and Midscene). Much of it
comes from apps whose tasks enter or edit many values (Markor, Recipe, Expense, Tasks; 42 tasks):
33.7 solved per trial for Coati against 29.3 for Artemis, 19.0 for FinalRun and 16.0 for Midscene.
Appendix~\ref{app:tasks} explains the tasks an opponent solved and Coati mostly did not.

\textbf{\textit{Published results.}}
Table~\ref{tab:androidworld-overall} in the appendix lists published and leaderboard results for
context. They use other backbones and harnesses, single trials, and are self-reported, so we do not
rank against them: a newer general model can do better than an older one without any change to the
agent. Coati lies within the range of recent published agents, most of which train their own model.

\textbf{\textit{From failures to fixes.}}
An earlier full trial reached 72.4\%. Most of its 32 failures were not reasoning errors but early
stops with budget left (14), step exhaustion after launches seen as blank screens (10) and a slider
that stopped short of its maximum (2). A same-target stopping rule, waiting for a launched app to
render and setting a slider to its extreme recovered 9 of them without losing any of 10 solved tasks.

\subsection{Reliability: Grounded Completion (RQ2)}
\label{subsec:rq2}

\begin{table}[t]
    \centering
    \small
    \caption{VerifyLoop on AndroidWorld (116 tasks, same build). C1: one trial; C2: mean of three. SR by
    difficulty and overall (\%). False done: share of declared completions that the evaluator judged
    failed (\%).}
    \label{tab:verify-ablation}
    \setlength{\tabcolsep}{4pt}
    \begin{tabular}{@{}lccccc@{}}
        \toprule
        \textbf{Configuration} & \textbf{Easy} & \textbf{Medium} & \textbf{Hard} & \textbf{SR} & \textbf{False done} \\
        \midrule
        C1: no VerifyLoop & 96.7 & 86.1 & 73.7 & 89.7 & 8.0 \\
        C2: Coati & 96.2 & 79.6 & 68.4 & 86.5 & 5.4 \\
        \bottomrule
    \end{tabular}
\end{table}

Table~\ref{tab:verify-ablation} isolates VerifyLoop. Without it, 9 of Coati's 112 declared
completions were false (8.0\%) and nothing caught them. With it, the audit rejected 61 completion
claims over the three trials (about twenty per trial); 40 of those tasks were then solved and 21
ended as honest reported failures, and only 16 of 299 accepted completions were false (5.4\%). The
false-completion rate thus fell from 8.0\% to 5.4\%, by about a third, at comparable success: C1
reaches 89.7\% in its single trial and C2 reaches the same 89.7\% in its matching first trial
(86.5\% averaged over three, the spread being trial-to-trial variance rather than a cost of the audit).

Host evidence matters most where a model cannot see the result. In a routed run of ``hide every
listing marked Sold'' before loop records existed, the fast executor hid one of six sold listings and
claimed completion; the remaining five were below the fold, and an audit of the final screen accepted
the claim. With loop delegation, the audit receives the loop record, including whether the loop
reached the end of the list, and the task succeeded in all runs (Table~\ref{tab:loops}). Conversely,
in classic mode one run of a loop task changed the app state correctly but was rejected three times
because no single final screen shows all six items: without host evidence, an honest auditor must
either accept unseen results or reject correct ones. Finally, the audit itself can fail: in the
earlier full trial, two of 116 completion audits timed out and the tasks were reported as done.
Coati now reports such completions as unverified (Section~\ref{subsec:verifyloop}).

\subsection{Speed in Everyday Use (RQ3)}
\label{subsec:rq3}


\begin{figure}[t]
\centering
\includegraphics [width=\linewidth]{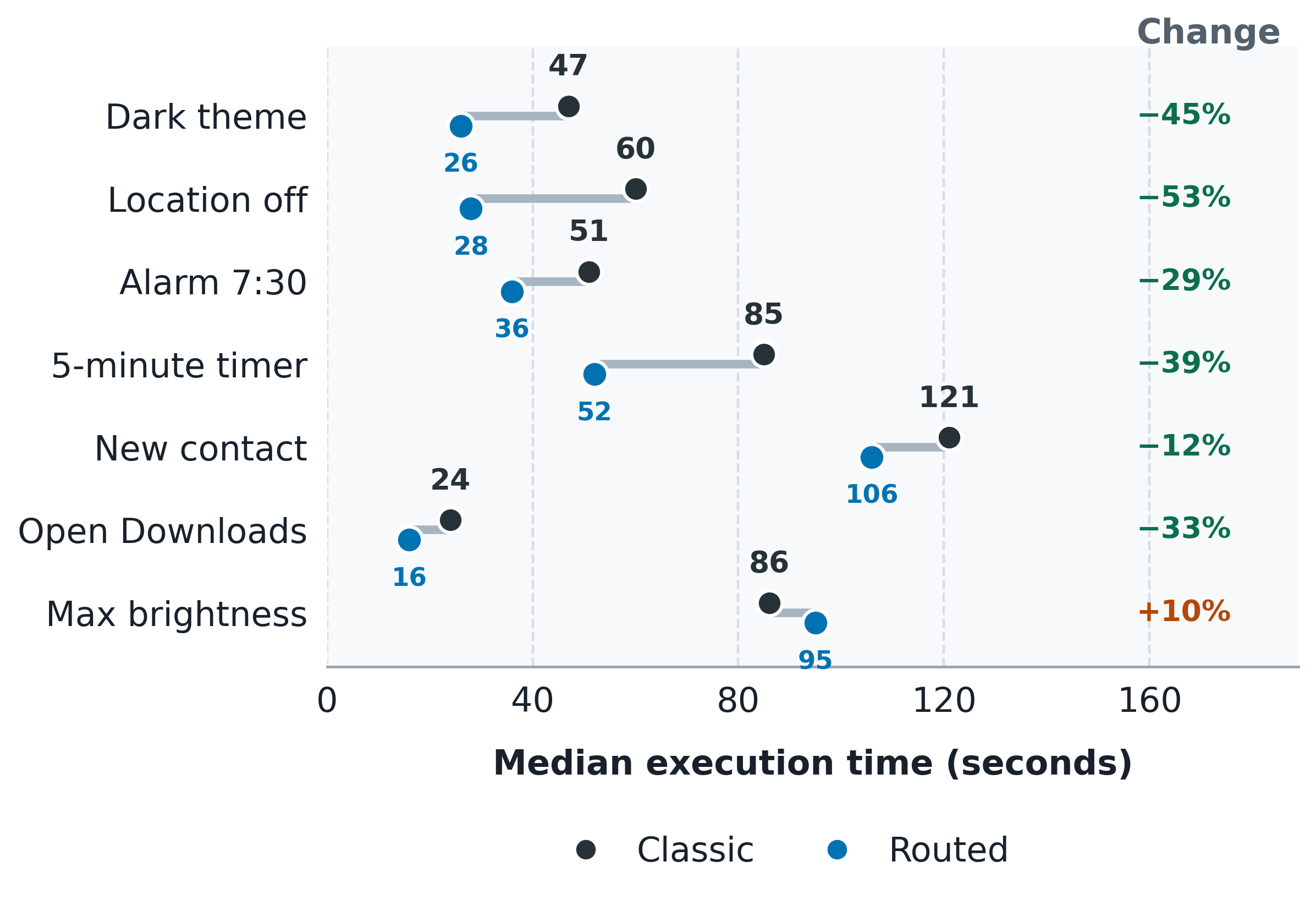}
\caption{Median execution time for seven tasks under Classic and Routed execution. Lower values indicate faster execution; annotations report successful runs and relative change.}
\label {fig:rq3-time}
\end {figure}

\begin {figure}[t]
\centering
\includegraphics [width=\linewidth]{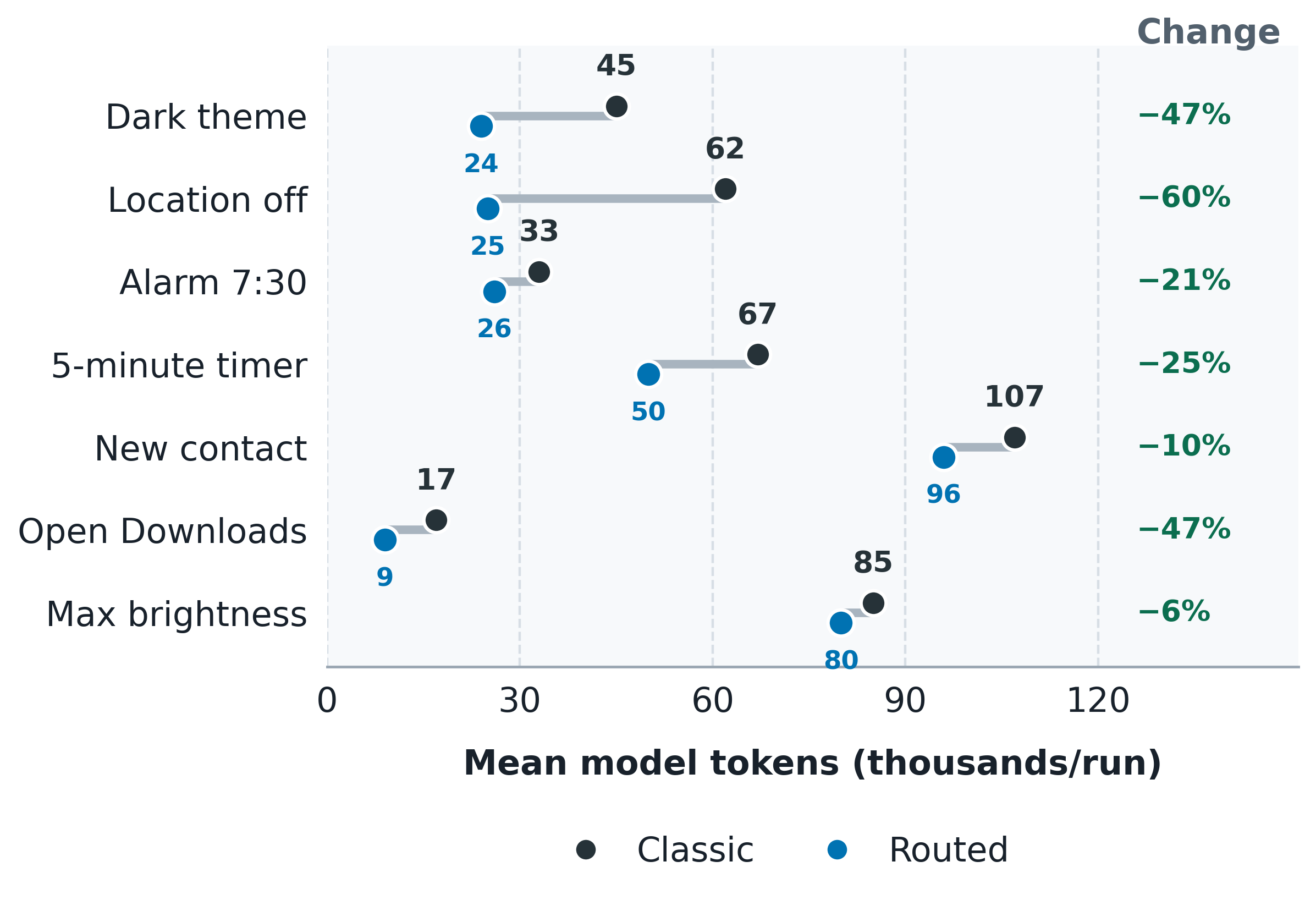}
\caption{Mean token usage (thousands per run) for seven tasks under Classic and Routed execution. Lower values indicate lower cost; annotations report successful runs and relative change.}
\label {fig:rq3-tokens}
\end {figure}

Figures~\ref{fig:rq3-time} and~\ref{fig:rq3-tokens} compare execution time and model-token usage on seven everyday tasks. The routed configuration is faster on six of them and cuts the median task time from 64\,s to 39\,s ($-39\%$) and mean token usage by 25\%, with comparable success. A fast step costs a tree parse and one selector round trip (about 2\,s) instead of a Reasoning call and a verification call; text entry and app switches still go to the Reasoning route, which is why the contact task gains least. The brightness slider remains the weak spot of both configurations. A camera task was excluded because its result check read the wrong directory for both.

\begin{table}[t]
    \centering
    \small
    \caption{Loops of simple steps in Car Market's repeat mode, five runs per configuration
    (R4 routed: three runs after the stop rule of Section~\ref{subsec:routing}). Time: mean seconds;
    steps: mean (fast-route steps in parentheses); tokens: mean thousands.}
    \label{tab:loops}
    \setlength{\tabcolsep}{1.8pt} 
    \begin{tabular}{@{}lcccccccc@{}}
        \toprule
        & \multicolumn{4}{c}{\textbf{Classic}} & \multicolumn{4}{c}{\textbf{Routed + loops}} \\
        \cmidrule(lr){2-5}\cmidrule(lr){6-9}
        \textbf{Task}
        & \thead{OK} & \thead{Time} & \thead{Steps} & \thead{Tok.}
        & \thead{OK} & \thead{Time} & \thead{Steps} & \thead{Tok.} \\
        \midrule
        R1 first 10 to favorites & 5/5 & 140 & 13 & 145 & 5/5 & \textbf{50} & 12 (11) & \textbf{21} \\
        R2 six given IDs to compare & 5/5 & 309 & 25 & 304 & 5/5 & \textbf{57} & 14 (13) & \textbf{33} \\
        R3 hide every Sold listing & 4/5 & 210 & 22 & 245 & 5/5 & \textbf{63} & 14 (13) & \textbf{39} \\
        R4 refresh until price drops & 5/5 & 100 & 9 & 101 & 3/3 & \textbf{92} & 8 (7) & \textbf{90} \\
        \bottomrule
    \end{tabular}
\end{table}

Loops of simple steps show the fast executor at its best (Table~\ref{tab:loops}). On R1--R3 the routed
configuration with loop delegation is 2.8--5.4$\times$ faster and uses 6--9$\times$ fewer tokens, at
equal or higher success: the Reasoning model plans the loop once, the selector finds the control, and
the host repeats it on every item without a model call. Without loop delegation the same selector
failed these tasks, scrolling past the first rows or stopping after one item. R4, where the condition
to stop is a price several rows below the refresh control, gains little: the stop condition is not
visible from the control, and the loop hands the task back to the Reasoning model, which checks it.
An earlier version kept scrolling in this case until the step limit; ending a loop after eight
actions without an item handled fixed it.

\paragraph{On a phone.} On an Android~16 phone with nine tasks over native, canvas, mixed,
permission-dialog and Flutter screens (Appendix~\ref{app:phone}), routed execution completes all 27
runs against 26 for per-step Reasoning, cuts the mean task time from 137.5\,s to 105.2\,s (median from 110.5\,s to 78.6\,s), is faster
on eight of the nine tasks, and reduces critical-path Reasoning calls from 285 to 67 (81\% of steps
decided on the fast route). Before a fast step taps a commit control such as \emph{Submit order}, a
one-call pre-commit gate checks the screen against the task; on the phone it stopped an order with
two copies instead of three (Appendix~\ref{subsec:case-precommit}).
Routed execution takes more steps (13.2 versus 10.4 on average) yet finishes sooner, because a fast
step costs a tree parse and one selector round trip instead of a Reasoning call and a verification
call. On the task whose target page lies several menus deep, both configurations explore without app
knowledge (31 and 34 steps); routed exploration is cheaper per step and completes every
run, and RQ4 shows that a learned manual removes this cost on repeated tasks. The exception is a
canvas task whose entry point lies below the first screen: the selector first searched the app for
the artwork title before scrolling, and each canvas step needs a vision parse.

The fast executor is not a general accelerator. AndroidWorld tasks are short (about thirteen steps on average), and
most steps there require reading the task and the screen together, which is the Reasoning route's
job; we therefore evaluate Coati's capability in its per-step configuration (RQ1) and the fast
executor where it is meant to help.

\subsection{Repeated Use: AppMemory (RQ4)}
\label{subsec:rq4}

\begin{table}[t]
    \centering
    \small
    \caption{AppMemory on a deep route in an unfamiliar app (Manual Lab, open the ``Archive'' page
    four menus deep) and on a familiar app (Messages, Advanced settings). Steps and seconds per run;
    each learning session starts from an empty manual.}
    \label{tab:appmemory}
    \setlength{\tabcolsep}{4pt}
    \begin{tabular}{@{}llcccc@{}}
        \toprule
        \textbf{App} & \textbf{Memory} & \textbf{Run 1} & \textbf{Run 2} & \textbf{Run 3} & \textbf{Run 4} \\
        \midrule
        Manual Lab & off & 14 / 171 & 19 / 198 & 18 / 183 & 19 / 198 \\
        & session 1 & 19 / 286 & \textbf{6 / 61} & \textbf{6 / 61} & \textbf{6 / 61} \\
        & session 2 & 14 / 138 & \textbf{7 / 80} & \textbf{7 / 71} & \textbf{7 / 85} \\
        & session 3 & 15 / 163 & \textbf{7 / 77} & \textbf{7 / 77} & \textbf{7 / 117} \\
        \midrule
        Messages & off & 5 / 46 & 5 / 49 & 5 / 52 & 5 / 46 \\
        & session 1 & 5 / 43 & 5 / 46 & 6 / 88 & 5 / 49 \\
        \bottomrule
    \end{tabular}
\end{table}

Table~\ref{tab:appmemory} shows three independent learning sessions on a deep route in an app the
model has never seen. Without memory, every run explores: 18.5 steps and 190\,s at the median. In each
session the first run explores and the manual learns the route from it; every later run goes there
directly, in 6--7 steps and a median of 77\,s, 2.5$\times$ faster, with all nine later runs
successful. The effect is specific to unfamiliar apps: on a familiar stock app the model already knows
the route, and memory neither helps nor hurts. The manual also survived an app update that moved the
entry points (version~2 of the fixture replaces two bottom tabs with a hub page): after learning on
version~1, runs on version~2 took 6 steps against 13--15 without memory.

The effect carries over to routed execution on the phone of RQ3: from an empty manual, the first
deep-route run explored for 22 steps (222\,s), and the next two took 5 steps (23\,s and 21\,s) against
34.3 steps and 295\,s without memory, with no Reasoning call on the critical path; all four runs, including a task on the same route, succeeded. On the fast route most
steps involve no model call that could describe the page; three changes made learning reliable there:
observation-only calls while the background advisor is busy, a per-task observation budget that
keeps the latest observations, where the route that worked is, rather than the first,
and a stricter page identity for learning (visible-text Jaccard $\ge 0.9$, not $0.6$), so that
sibling pages sharing their bars are described separately.

\section{Lessons from Deployment}
\label{sec:lessons}

Deploying the agent as an Android app exposed failure modes that affect both success and speed. Accessibility-based agents are likely to encounter the same issues.
\emph{The agent's own overlay hides the app.} Android marks every node under a touchable window as
invisible to the user; a tree read while Coati's overlay was touchable kept 2 of 63 nodes, so the fast
route saw an almost empty screen. Reading the tree while the overlay passes touches restored 62 of 63.
\emph{Injected taps are not always taps.} Accessibility gestures carry no touch source, and some list
views ignore them; Coati falls back to the node's click action when a list item's tap leaves the
screen unchanged. Presses of 50\,ms or longer were dropped by one app right after launch, so taps last
20\,ms, and swipes are kept out of the system gesture areas, where they trigger navigation.
\emph{State must survive power loss.} Task state written without a file sync was empty after a cold
restart; writes now sync before an atomic rename.
\emph{Measurement interferes.} Using \texttt{uiautomator} to check results suspends the agent's
accessibility service; checks read state through the agent instead.
\emph{Coordinate spaces and thinking budgets.} Mixing pixel bounds with normalized coordinates in one
prompt made the model copy pixel values, and output limits that include thinking tokens truncated
answers; both were fixed in the prompt and request settings.

\section{Discussion and Limitations}
\label{sec:discussion}

\textbf{\textit{Scope and generalization.}}
AndroidWorld evaluates first-time task completion, but not routine execution time,
repeated use, or the reliability of completion claims. These aspects are evaluated
separately in RQ3 and RQ4. Routing is less effective during first-time exploration,
especially on deep routes, whereas AppMemory reduces this cost on later visits.
Our main experiments use emulators, one backbone model, two released fixture apps,
and a limited number of runs; the results also depend on server-side model latency.
The phone results in Appendix~\ref{app:phone} provide an additional check on current
consumer hardware, but broader evaluations across devices, models and task sets remain
future work.

\textbf{\textit{Hardware and privacy.}}
GPU-assisted OCR requires a usable Vulkan GPU. Without one, Coati falls back to
the accessibility tree and vision parsing, and the OCR models also increase the
application size. Screenshots and page descriptions are sent to model providers.
Although AppMemory stores only sanitized, interface-supported facts, neither
mechanism prevents sensitive data from leaving the device; deployments should
therefore use trusted model endpoints.

\textbf{\textit{Deployment and platform policy.}}
Coati operates other applications through Android's Accessibility Service, whose
autonomous use is restricted by current Google Play policy
~\cite{gplay-a11y-policy,gplay-policy-2025}. We therefore distribute Coati as
source code for direct installation rather than through Google Play. The proposed
verification, memory and routing mechanisms are independent of the action-injection
layer and could transfer to sanctioned system interfaces. In the current design,
permission and installation dialogs remain under user control, actions are visible
on screen, and users can pause or stop a task at any time.

\newpage
\bibliographystyle{ACM-Reference-Format}
\balance
\bibliography{refs}

@misc{appium,
    title        = {Appium: Cross-Platform Automation Framework for Mobile and Desktop Apps},
    author       = {{Appium Contributors}},
    year         = {2026},
    howpublished = {\url{https://appium.io/}},
    note         = {Open-source UI automation framework}
}

@inproceedings{cheng-etal-2024-seeclick,
    title     = {{SeeClick}: Harnessing {GUI} Grounding for Advanced Visual {GUI} Agents},
    author    = {Cheng, Kanzhi and Sun, Qiushi and Chu, Yougang and Xu, Fangzhi and YanTao, Li and Zhang, Jianbing and Wu, Zhiyong},
    booktitle = {Proceedings of the 62nd Annual Meeting of the Association for Computational Linguistics},
    pages     = {9313--9332},
    year      = {2024},
    address   = {Bangkok, Thailand},
    publisher = {Association for Computational Linguistics},
    doi       = {10.18653/v1/2024.acl-long.505}
}

@article{qin2025uitars,
    title   = {{UI-TARS}: Pioneering Automated {GUI} Interaction with Native Agents},
    author  = {Qin, Yujia and Ye, Yining and Fang, Junjie and others},
    journal = {arXiv preprint arXiv:2501.12326},
    year    = {2025}
}

@inproceedings{lai-etal-2025-androidgen,
    title     = {{AndroidGen}: Building an Android Language Agent under Data Scarcity},
    author    = {Lai, Hanyu and Gao, Junjie and Liu, Xiao and others},
    booktitle = {Proceedings of the 63rd Annual Meeting of the Association for Computational Linguistics},
    pages     = {2727--2749},
    year      = {2025},
    address   = {Vienna, Austria},
    publisher = {Association for Computational Linguistics},
    doi       = {10.18653/v1/2025.acl-long.138}
}

@inproceedings{wang2024mobileagent,
    title     = {Mobile-Agent: Autonomous Multi-Modal Mobile Device Agent with Visual Perception},
    author    = {Wang, Junyang and Xu, Haiyang and Ye, Jiabo and others},
    booktitle = {ICLR 2024 Workshop on Large Language Model Agents},
    year      = {2024}
}

@inproceedings{wang2024mobileagentv2,
    title     = {Mobile-Agent-v2: Mobile Device Operation Assistant with Effective Navigation via Multi-Agent Collaboration},
    author    = {Wang, Junyang and Xu, Haiyang and Jia, Haitao and others},
    booktitle = {Advances in Neural Information Processing Systems},
    volume    = {37},
    year      = {2024},
    url       = {https://proceedings.neurips.cc/paper_files/paper/2024/hash/0520537ba799d375b8ff5523295c337a-Abstract-Conference.html}
}

@inproceedings{rawles2025androidworld,
    title     = {AndroidWorld: A Dynamic Benchmarking Environment for Autonomous Agents},
    author    = {Rawles, Christopher and Clinckemaillie, Sarah and Chang, Yifan and others},
    booktitle = {International Conference on Learning Representations},
    year      = {2025}
}

@inproceedings{wen2024autodroid,
    title     = {{AutoDroid}: {LLM}-powered Task Automation in {Android}},
    author    = {Wen, Hao and Li, Yuanchun and Liu, Guohong and others},
    booktitle = {Proceedings of the 30th Annual International Conference on Mobile Computing and Networking},
    pages     = {543--557},
    year      = {2024},
    publisher = {ACM},
    doi       = {10.1145/3636534.3649379}
}

@article{li2024appagentv2,
    title   = {AppAgent v2: Advanced Agent for Flexible Mobile Interactions},
    author  = {Li, Yanda and Zhang, Chi and Jiang, Wenjia and others},
    journal = {arXiv preprint arXiv:2408.11824},
    year    = {2024}
}

@inproceedings{xie2025guiexplorer,
    title     = {{GUI}-explorer: Autonomous Exploration and Mining of Transition-aware Knowledge for {GUI} Agent},
    author    = {Xie, Bin and Shao, Rui and Chen, Gongwei and others},
    booktitle = {Proceedings of the 63rd Annual Meeting of the Association for Computational Linguistics},
    pages     = {5650--5667},
    year      = {2025},
    address   = {Vienna, Austria},
    publisher = {Association for Computational Linguistics},
    doi       = {10.18653/v1/2025.acl-long.282}
}

@inproceedings{zhao2025llmexplorer,
    title     = {LLM-Explorer: Towards Efficient and Affordable LLM-based Exploration for Mobile Apps},
    author    = {Zhao, Shanhui and Wen, Hao and Du, Wenjie and others},
    booktitle = {Proceedings of the 31st Annual International Conference on Mobile Computing and Networking},
    year      = {2025},
    publisher = {ACM},
    note      = {arXiv:2505.10593}
}

@inproceedings{li2026curiosity,
    title     = {Curiosity Driven Knowledge Retrieval for Mobile Agents},
    author    = {Li, Sijia and Tan, Xiaoyu and Ali, Shahir and others},
    booktitle = {Proceedings of the ACM Web Conference 2026},
    year      = {2026},
    publisher = {ACM},
    doi       = {10.1145/3774904.3792257}
}

@inproceedings{li2025mobileuse,
    title     = {MobileUse: A Hierarchical Reflection-Driven {GUI} Agent for Autonomous Mobile Operation},
    author    = {Li, Ning and Qu, Xiangmou and Zhou, Jiamu and others},
    booktitle = {Advances in Neural Information Processing Systems},
    volume    = {38},
    pages     = {40361--40388},
    year      = {2025},
    publisher = {Curran Associates, Inc.},
    doi       = {10.52202/085713-1347}
}

@inproceedings{lee2025verisafe,
    title     = {VeriSafe Agent: Safeguarding Mobile GUI Agent via Logic-based Action Verification},
    author    = {Lee, Jungjae and Lee, Dongjae and Choi, Chihun and others},
    booktitle = {Proceedings of the 31st Annual International Conference on Mobile Computing and Networking},
    pages     = {817--831},
    year      = {2025},
    address   = {Hong Kong, China},
    publisher = {ACM},
    doi       = {10.1145/3680207.3765248}
}

@inproceedings{dai2026vdroid,
    title     = {V-Droid: Advancing Mobile {GUI} Agent Through Generative Verifiers},
    author    = {Dai, Gaole and Jiang, Shiqi and Cao, Ting and others},
    booktitle = {Proceedings of the 32nd Annual International Conference on Mobile Computing and Networking},
    year      = {2026},
    publisher = {ACM}
}

@inproceedings{hong2024cogagent,
  title     = {{CogAgent}: A Visual Language Model for {GUI} Agents},
  author    = {Hong, Wenyi and Wang, Weihan and Lv, Qingsong and others},
  booktitle = {Proceedings of the IEEE/CVF Conference on Computer Vision
               and Pattern Recognition},
  pages     = {14281--14290},
  year      = {2024}
}

@inproceedings{zheng2024seeact,
  title     = {{GPT-4V(ision)} is a Generalist Web Agent, if Grounded},
  author    = {Zheng, Boyuan and Gou, Boyu and Kil, Jihyung
               and others},
  booktitle = {Proceedings of the 41st International Conference
               on Machine Learning},
  series    = {Proceedings of Machine Learning Research},
  volume    = {235},
  pages     = {61349--61385},
  year      = {2024},
  url       = {https://proceedings.mlr.press/v235/zheng24e.html}
}

@inproceedings{nguyen2025guiagents,
  title     = {{GUI} Agents: A Survey},
  author    = {Nguyen, Dang and Chen, Jian and Wang, Yu and others},
  booktitle = {Findings of the Association for Computational Linguistics:
               ACL 2025},
  pages     = {22522--22538},
  year      = {2025},
  address   = {Vienna, Austria},
  publisher = {Association for Computational Linguistics},
  doi       = {10.18653/v1/2025.findings-acl.1158},
  url       = {https://aclanthology.org/2025.findings-acl.1158/}
}

@inproceedings{rawles2023androidwild,
  title     = {{AndroidInTheWild}: A Large-Scale Dataset For Android Device Control},
  author    = {Rawles, Christopher and Li, Alice and Rodriguez, Daniel
               and others},
  booktitle = {Advances in Neural Information Processing Systems
               (Datasets and Benchmarks Track)},
  volume    = {36},
  pages     = {59708--59728},
  year      = {2023},
  doi       = {10.52202/075280-2609},
  url       = {https://proceedings.neurips.cc/paper_files/paper/2023/hash/bbbb6308b402fe909c39dd29950c32e0-Abstract-Datasets_and_Benchmarks.html}
}

@inproceedings{zhang2025appagent,
  title     = {{AppAgent}: Multimodal Agents as Smartphone Users},
  author    = {Zhang, Chi and Yang, Zhao and Liu, Jiaxuan
               and others},
  booktitle = {Proceedings of the 2025 CHI Conference
               on Human Factors in Computing Systems},
  series    = {CHI '25},
  year      = {2025},
  publisher = {Association for Computing Machinery},
  doi       = {10.1145/3706598.3713600}
}

@inproceedings{wu2025backtrackagent,
  title     = {{BacktrackAgent}: Enhancing {GUI} Agent with Error Detection
               and Backtracking Mechanism},
  author    = {Wu, Qinzhuo and Gao, Pengzhi and Liu, Wei and Luan, Jian},
  booktitle = {Proceedings of the 2025 Conference on Empirical Methods
               in Natural Language Processing},
  pages     = {4250--4272},
  year      = {2025},
  address   = {Suzhou, China},
  publisher = {Association for Computational Linguistics},
  doi       = {10.18653/v1/2025.emnlp-main.212},
  url       = {https://aclanthology.org/2025.emnlp-main.212/}
}

@inproceedings{zhang2026verigui,
  title     = {Don{'}t Act Blindly: Robust {GUI} Automation via Action-Effect
               Verification and Self-Correction},
  author    = {Zhang, Yuzhe and Xue, Xianwei and Wu, Xingyong and others},
  booktitle = {Proceedings of the 64th Annual Meeting of the Association
               for Computational Linguistics (Volume 1: Long Papers)},
  pages     = {28924--28941},
  year      = {2026},
  publisher = {Association for Computational Linguistics},
  doi       = {10.18653/v1/2026.acl-long.1335},
  url       = {https://aclanthology.org/2026.acl-long.1335/}
}

@inproceedings{yang-etal-2025-aria,
    title = "Aria-{UI}: Visual Grounding for {GUI} Instructions",
    author = "Yang, Yuhao  and
      Wang, Yue  and
      Li, Dongxu  and others",
    booktitle = "Findings of the Association for Computational Linguistics: ACL 2025",
    month = jul,
    year = "2025",
    address = "Vienna, Austria",
    publisher = "Association for Computational Linguistics",
    url = "https://aclanthology.org/2025.findings-acl.1152/",
    doi = "10.18653/v1/2025.findings-acl.1152",
    pages = "22418--22433",
    ISBN = "979-8-89176-256-5"
}

@inproceedings{agashe2025agents2,
  title     = {Agent S2: A Compositional Generalist-Specialist Framework for Computer Use Agents},
  author    = {Agashe, Saaket and Wong, Kyle and Tu, Vincent and others},
  booktitle = {Second Conference on Language Modeling},
  year      = {2025},
  url       = {https://openreview.net/forum?id=zg5is4GJ3R}
}

@misc{ye2025mobileagentv3,
      title={Mobile-Agent-v3: Fundamental Agents for GUI Automation}, 
      author={Jiabo Ye and Xi Zhang and Haiyang Xu and others},
      year={2025},
      eprint={2508.15144},
      archivePrefix={arXiv},
      primaryClass={cs.AI},
      url={https://arxiv.org/abs/2508.15144}, 
}

@misc{yan2026openmobile2,
  title         = {{OpenMobile-2}: Building Versatile Mobile Agents with
                   Scalable Environments and App-Native Tools},
  author        = {Yan, Chenyang and Zhang, Yingying and Cheng, Kanzhi
                   and Sun, Qiushi and others},
  year          = {2026},
  eprint        = {2610.14157},
  archivePrefix = {arXiv},
  primaryClass  = {cs.SE},
  url           = {https://arxiv.org/abs/2610.14157}
}

@inproceedings{xu2026mobilerl,
  title     = {{MobileRL}: Online Agentic Reinforcement Learning
               for Mobile GUI Agents},
  author    = {Xu, Yifan and Liu, Xiao and Liu, Xinghan
               and others},
  booktitle = {The Fourteenth International Conference
               on Learning Representations},
  year      = {2026},
  url       = {https://openreview.net/forum?id=C3F0G9nXhl}
}

@misc{androidworldleaderboard,
  author       = {{Google Research}},
  title        = {{AndroidWorld} Leaderboard},
  year         = {2026},
  howpublished = {\url{https://docs.google.com/spreadsheets/d/1cchzP9dlTZ3WXQTfYNhh3avxoLipqHN75v1Tb86uhHo}},
  note         = {Linked from \url{https://github.com/google-research/android_world}. Accessed October 1, 2026}
}

@article{lu2024omniparser,
  title     = {{OmniParser} for Pure Vision Based {GUI} Agent},
  author    = {Lu, Yadong and Yang, Jianwei and Shen, Yelong and Awadallah, Ahmed},
  journal   = {arXiv preprint arXiv:2408.00203},
  year      = {2024}
}

@article{du2020ppocr,
  title     = {{PP-OCR}: A Practical Ultra Lightweight {OCR} System},
  author    = {Du, Yuning and Li, Chenxia and Guo, Ruoyu and others},
  journal   = {arXiv preprint arXiv:2009.09941},
  year      = {2020}
}

@article{cui2025paddleocr3,
  title     = {{PaddleOCR} 3.0 Technical Report},
  author    = {Cui, Cheng and Sun, Ting and Lin, Manhui and others},
  journal   = {arXiv preprint arXiv:2507.05595},
  year      = {2025}
}

@misc{ncnn,
  title        = {ncnn: A High-Performance Neural Network Inference Framework Optimized for the
                  Mobile Platform},
  author       = {{Tencent}},
  howpublished = {\url{https://github.com/Tencent/ncnn}},
  year         = {2017}
}

@inproceedings{lin2023swiftsage,
  title     = {{SwiftSage}: A Generative Agent with Fast and Slow Thinking for Complex
               Interactive Tasks},
  author    = {Lin, Bill Yuchen and Fu, Yicheng and Yang, Karina and others},
  booktitle = {Advances in Neural Information Processing Systems},
  year      = {2023}
}

@article{chen2023frugalgpt,
  title     = {{FrugalGPT}: How to Use Large Language Models While Reducing Cost and
               Improving Performance},
  author    = {Chen, Lingjiao and Zaharia, Matei and Zou, James},
  journal   = {Transactions on Machine Learning Research},
  year      = {2024}
}

@inproceedings{ong2024routellm,
  title     = {{RouteLLM}: Learning to Route {LLMs} from Preference Data},
  author    = {Ong, Isaac and Almahairi, Amjad and Wu, Vincent and others},
  booktitle = {The Thirteenth International Conference on Learning Representations},
  year      = {2025}
}

@misc{typesafejev,
  title        = {Jev: A System One Model for Typed Decisions},
  author       = {{TypeSafe AI}},
  howpublished = {\url{https://docs.typesafe.ai/}},
  note         = {Model version jev-1.13.0},
  year         = {2026}
}

@misc{artemis2026,
  author       = {{Google}},
  title        = {Artemis},
  year         = {2026},
  howpublished = {GitHub repository},
  url          = {https://github.com/google/artemis},
  note         = {Accessed October 1, 2026}
}

@misc{finalrun2026,
  author       = {{Finalrun}},
  title        = {Finalrun Agent},
  year         = {2026},
  howpublished = {GitHub repository},
  url          = {https://github.com/final-run/finalrun-agent},
  note         = {Accessed October 1, 2026}
}

@misc{midscene2026,
  author       = {{Midscene.js Team}},
  title        = {{Midscene.js} {AndroidWorld} Benchmark Report},
  year         = {2026},
  howpublished = {\url{https://midscenejs.com/android-world-benchmark-report}},
  note         = {Accessed October 1, 2026}
}

@misc{gplay-a11y-policy,
  author       = {{Google}},
  title        = {Use of the {AccessibilityService} {API}},
  howpublished = {Play Console Help},
  year         = {2026},
  url          = {https://support.google.com/googleplay/android-developer/answer/10964491},
  note         = {Accessed 2026-10-18}
}

@misc{gplay-policy-2025,
  author       = {{Google}},
  title        = {Policy announcement: {October} 30, 2025},
  howpublished = {Play Console Help},
  year         = {2025},
  url          = {https://support.google.com/googleplay/android-developer/answer/16550159},
  note         = {Accessed 2026-10-18}
}

\appendix

\section{Perception Details}
\label{app:perception}

\paragraph{On-device GPU OCR.}
Coati runs PP-OCRv5 mobile~\cite{du2020ppocr,cui2025paddleocr3} through NCNN with Vulkan~\cite{ncnn} on the device GPU.
Devices without a usable Vulkan GPU run without OCR: a CPU implementation took 1.5--6\,s per screen on
an entry-level device, too slow to help a step. On a recent phone GPU, a complete OCR pass takes 40--85\,ms
per screen in the runs of Appendix~\ref{app:phone}; on an entry-level Mali-G52, detection takes about
0.2\,s and recognition about 90\,ms per line, and batching horizontally adjacent lines into one recognition pass
reduces total time by 25--35\% at unchanged accuracy (an 84-line screen from 7.8\,s to 5.9\,s, a
21-line screen from 2.0\,s to 1.5\,s). Recognition skips labelled nodes, the status bar and the soft
keyboard, and on native screens reads only unlabelled, non-image nodes. Models load asynchronously at
service start, and on tree-less screens OCR runs in parallel with the vision parse and may delay a
step by at most one second.

\paragraph{Conservative OCR repair.}
A first repair strategy that replaced model text with OCR text improved parses overall but degraded
elements the model had right: OCR misreads small text, icon glyphs and thumbnails. Coati's final
repair never overwrites model text. It keeps only confident lines (confidence $\ge 0.8$ with at least
two alphanumeric characters, or confident single-digit counters), appends confident lines the parse
missed, fills elements that have no text, snaps text-like elements only to lines with the same
content, and moves an element only when the match is unique. Section~\ref{subsec:exp-perception}
shows that this strategy improves every metric without degrading any element.

\paragraph{Parse agreement.}
If fewer than 30\% of a vision parse's texts are confirmed by OCR on the same screenshot, the parse
describes a different image and is requested again. We observed such parses through a third-party
model relay that mixed up responses between requests; the check discards them before they cause
actions, although it cannot prevent the screenshot from having been sent.

\paragraph{\textbf{Parse-level perception benchmark.}}
Fourteen device screenshots (four native, three custom canvas, three Flutter, two from a video app, two
from a streaming client), 28 vision parses, with ground-truth element texts and positions from the
accessibility tree or layout constants. We report the share of ground-truth elements found, the share
with exactly correct text, the share whose parsed tap point falls inside the true element, and the
number of elements that a repair made worse.

\subsection{Parse-Level Evaluation}
\label{subsec:exp-perception}

\begin{table}[t]
    \centering
    \small
    \caption{Parse-level evaluation on 14 device screens (28 vision parses). Worse counts elements
    whose text or tap position a repair made worse.}
    \label{tab:ocr-parse}
    \begin{tabular}{@{}lccc@{}}
        \toprule
        \textbf{Metric} & \textbf{Vision parse} & \textbf{+ OCR repair} & $\Delta$ \\
        \midrule
        Elements found & 51.7\% & \textbf{63.7\%} & +12.0 \\
        Exact text & 49.2\% & \textbf{57.6\%} & +8.4 \\
        Tap hit & 26.4\% & \textbf{32.0\%} & +5.6 \\
        Elements made worse & -- & \textbf{0} & \\
        \bottomrule
    \end{tabular}
\end{table}

Table~\ref{tab:ocr-parse} shows that conservative OCR repair improves every parse-level metric
without degrading a single element; the earlier replacing strategy improved the totals as well but
corrupted correct elements (``1.3M subscribers'' became ``3M sul'', ``Background'' became
``F Backgr''), which motivated never overwriting model text. On an offline selection set of 96
annotated phrases, a bundled mobile OCR model matched 77\% (below the vision model's own 91\%), while
PP-OCR matched all 96, which led to the NCNN deployment. On native screens with a usable tree, OCR
reads only unlabelled nodes and typically completes in 0.4--0.6\,s, and the hybrid trigger flags
only screens that mix native and painted content.

Tap-hit remains the weakest metric: vision parses place elements imprecisely on custom-drawn screens,
which is why Coati prefers tree bounds wherever the tree is usable and executes fast-route actions at
the element's own bounds.

\section{Evaluation Protocol}
\label{app:protocol}

\paragraph{Harness.}
All agents run through one harness that instantiates AndroidWorld tasks from the official registry
(seed 30), applies the official step budgets, records every attempt in a per-emulator ledger, cold
restarts the emulator every ten attempts, and scores with the official evaluators. The harness
refuses to resume an experiment whose identity (task instance, parameters, sources and model
configuration) differs from the stored one, so every reported episode ran under one configuration.

\paragraph{Environment failures and resumption.}
An episode is excluded and rerun, for every agent alike, only when the environment failed before the
agent could act: AndroidWorld's own task setup raised an error (for example a missing tool or an
uninitialized database), the agent's driver did not start, or no model response arrived at all. One
task, MarkorCreateFolder, names its folder after the current time to the second, so its instance can
never match a stored one when a run resumes; its stored episode is rerun on every resumption, again
for every agent. Archived episodes are kept with the results. We release the harness, its settings and every per-run result.

\paragraph{Published results.}
Table~\ref{tab:androidworld-overall} lists published and leaderboard results for context.

\begin{table*}[t]
    \centering
    \small
    \caption{Success on AndroidWorld. SR is the task success rate (\%). (a) Published methods, with
    numbers from their papers. (b) Submissions to the public AndroidWorld
    leaderboard~\cite{androidworldleaderboard} (accessed October 2026), which are self-reported,
    single-trial and not independently verified; we summarize them rather than rank against them.
    (c) Coati under the protocol of Section~\ref{subsec:experimental-settings}, mean over three
    trials. A11y tree denotes the Android accessibility UI tree.}
    \label{tab:androidworld-overall}
    \renewcommand{\arraystretch}{1.08}
    \setlength{\tabcolsep}{7pt}
    \begin{tabular}{@{}lp{0.34\textwidth}p{0.23\textwidth}c@{}}
        \toprule
        \textbf{Method} & \textbf{Model} & \textbf{Screen Representation} & \textbf{SR} \\
        \midrule
        \multicolumn{4}{@{}l}{\emph{(a) Published methods (paper-reported)}} \\
        UI-TARS~\cite{qin2025uitars} & UI-TARS-72B & Screenshot & 46.6 \\
        Agent S2~\cite{agashe2025agents2} & Agent S2 & Screenshot & 54.3 \\
        V-Droid~\cite{dai2026vdroid} & V-Droid (Llama-8B) & A11y tree & 59.5 \\
        Mobile-Agent-v3~\cite{ye2025mobileagentv3} & GUI-Owl-32B & Screenshot & 73.3 \\
        MobileRL~\cite{xu2026mobilerl} & MobileRL-9B (GLM-4.1V-9B) & Screenshot + A11y tree & 80.2 \\
        \midrule
        \multicolumn{4}{@{}l}{\emph{(b) Leaderboard submissions (self-reported, single trial, not independently verified)}} \\
        Top seven submissions & Mostly proprietary models (e.g., GPT-5.4 to GPT-6, Gemini~3.7,
        Seed1.8) & Screenshot, or Screenshot + A11y tree & 97.4--100 \\
        Human reference~\cite{rawles2025androidworld} & -- & -- & 80.0 \\
        \midrule
        \multicolumn{4}{@{}l}{\emph{(c) This work (three trials)}} \\
        \textbf{Coati (Ours)} & Gemini~3.8~Flash & \textbf{Screenshot + A11y tree} & 86.5 \\
        \bottomrule
    \end{tabular}
\end{table*}

\section{Tasks}
\label{app:tasks}

\paragraph{Everyday tasks.}
Turn on dark theme; turn off location; set an alarm for 7:30 AM in the Clock app; start a 5-minute
timer in the Clock app; create a contact named Alex Chen with phone number 555 123 4567; open the
Downloads folder in the Files app; set the screen brightness to the maximum. Results are checked
through system settings, the contacts provider, or the final screen.

\paragraph{Loops of simple steps.}
Car Market is an offline marketplace fixture with 40 listings. In its repeat mode every row shows the
listing ID, a Sold badge on six listings, and Favorite, Compare and Hide controls; a Refresh control
drops one listing's price on the fifth refresh. R1: add the first 10 listings to favorites. R2: add
six given listing IDs to the compare list. R3: hide every listing marked Sold. R4: refresh until the
price of a named listing drops below \$40{,}000 and report it. The fixture persists its state, which
the checker reads after each run.

\paragraph{Repeated use.}
Manual Lab is an offline fixture with four bottom tabs and a low-frequency route
(Me~$\rightarrow$~Settings~$\rightarrow$~Advanced settings~$\rightarrow$~Archive); its version~2
keeps the package identity and replaces two tabs with a tool hub. The familiar-app task opens the
Advanced settings of Google Messages. Both fixtures, all task definitions and checkers are released
with the code.

\paragraph{AndroidWorld tasks lost to an opponent.}
Seven tasks were solved by FinalRun or Midscene in their single trial while Coati averaged below one
half over three trials. \emph{CameraTakePhoto} (1 of 3): in two trials the camera's shutter did not
respond to Coati's taps, and the audit accepted the completion because a viewfinder looks the same
before and after a photo. \emph{SimpleCalendarNextMeetingWithPerson} (1 of 3): Coati found the
right event but ended on its date picker, which shows the date and hides the time, and twice
reported a wrong hour. \emph{SportsTrackerActivitiesOnDate} (1 of 3): twice Coati named one of the
two activity types recorded on the date. \emph{SystemBrightnessMaxVerify} (0 of 3): a tap at the
end of the brightness slider fills it visibly but leaves the setting short of the maximum, which no
screen shows. \emph{TasksHighPriorityTasksDueOnDate} (1 of 3): Coati opened tasks one by one to
read their priority and due date, each closing through a discard dialog, and twice ran out of its
ten steps. \emph{VlcCreatePlaylist} (1 of 3) and \emph{VlcCreateTwoPlaylists} (1 of 3): selecting
several files in order in VLC's file picker exhausted the step budget in two trials each. Three
of the seven tasks thus ran out of steps; the other four ended in completions whose error the final
screens do not show, the limit of a screen-based audit discussed in Section~\ref{subsec:rq2}.

\section{Routed Execution on a Phone}
\label{app:phone}

\paragraph{\textbf{On-device benchmark.}}
To measure routed execution on current consumer hardware, we built a benchmark of nine tasks on a
OnePlus phone running Android~16 (1272$\times$2772 portrait). The tasks cover interface types that
stress different components (Table~\ref{tab:device-bench}): native pages reached through several menus
(L1), search with text entry across pages (L2), a setting that depends on another page (L3), a
custom-drawn canvas with no accessibility tree (W1) and with text entry (W2), a native form with
painted stepper and toggle widgets (O1), a system permission dialog shown by another process (P1), and
two Flutter apps: a finance app with login and tab navigation (G1) and settings (G2). Each task has an
automatic check of the resulting app state. Both configurations use Gemini~3.8~Flash as Reasoning and
Fast model through the official API; the fast route's selector is Jev (version jev-1.13.0), a
``System One'' model that answers structured questions with typed, confidence-scored
decisions~\cite{typesafejev}. \emph{Per-step Reasoning} is Coati with every step on the Reasoning route
(VerifyLoop enabled); \emph{Routed} enables the fast route in \textsc{Fast} mode, hybrid perception and
GPU OCR. Each task runs three times per configuration, both configurations in every round, with app
knowledge off. Task time is the duration the agent logs from task start to accepted completion. Model
latency in the test environment fluctuated, so absolute times are longer than in the emulator
experiments and vary between rounds; both configurations are affected alike.

\subsection{Routed versus Per-step Execution on a Phone}
\label{subsec:exp-routing}

\begin{table*}[t]
    \centering
    \small
    \caption{On-device benchmark (Android~16 phone, Gemini 3.8 Flash for Reasoning and Fast). Each
    cell reports verified successes over three runs, mean task time of the verified runs, mean steps,
    and mean Reasoning calls on the critical path per run (the routed configuration's background
    planning calls are not counted).}
    \label{tab:device-bench}
    \renewcommand{\arraystretch}{1.08}
    \setlength{\tabcolsep}{4.5pt}
    \begin{tabular}{@{}llcccccccc@{}}
        \toprule
        & & \multicolumn{4}{c}{\textbf{Per-step Reasoning}} & \multicolumn{4}{c}{\textbf{Routed (Coati)}} \\
        \cmidrule(lr){3-6}\cmidrule(lr){7-10}
        \textbf{Task} & \textbf{Interface} & \textbf{Succ.} & \textbf{Time (s)} & \textbf{Steps} & \textbf{Reas.}
        & \textbf{Succ.} & \textbf{Time (s)} & \textbf{Steps} & \textbf{Reas.} \\
        \midrule
        L1 & native, several menus deep & 2/3 & 410.9 & 31.0 & 31.0 & \textbf{3/3} & \textbf{295.2} & 34.3 & 13.7 \\
        L2 & native, search and text entry & 3/3 & 78.7 & 6.0 & 6.0 & 3/3 & \textbf{32.4} & 6.3 & 0.0 \\
        L3 & native, dependent setting & 3/3 & 60.9 & 5.0 & 5.0 & 3/3 & \textbf{21.2} & 5.0 & 0.0 \\
        W1 & custom canvas, no tree & 3/3 & 99.1 & 7.0 & 7.0 & 3/3 & \textbf{82.7} & 9.0 & 0.7 \\
        W2 & canvas and text entry & 3/3 & \textbf{100.5} & 7.7 & 7.7 & 3/3 & 120.8 & 16.3 & 1.0 \\
        O1 & native form with painted widgets & 3/3 & 238.1 & 15.7 & 15.7 & 3/3 & \textbf{191.9} & 19.0 & 2.7 \\
        P1 & system permission dialog & 3/3 & 159.5 & 9.7 & 10.7 & 3/3 & \textbf{71.5} & 11.7 & 2.0 \\
        G1 & Flutter, login and tabs & 3/3 & 128.7 & 8.0 & 8.0 & 3/3 & \textbf{85.5} & 10.3 & 1.3 \\
        G2 & Flutter, settings & 3/3 & 51.8 & 4.0 & 4.0 & 3/3 & \textbf{45.5} & 7.0 & 1.0 \\
        \midrule
        \multicolumn{2}{@{}l}{\textbf{All (27 runs)}} & 26/27 & 137.5 & 10.4 & 285 total & \textbf{27/27} & \textbf{105.2} & 13.2 & \textbf{67 total} \\
        \bottomrule
    \end{tabular}
\end{table*}

Table~\ref{tab:device-bench} compares the two configurations; Section~\ref{subsec:rq3} summarizes
the totals. Per-step Reasoning's one failure is an L1 run that exhausted the time limit while it
searched for the deep page. Routed execution is faster by 55--65\% on native search, the dependent
setting and the permission dialog and by 34\% on the Flutter login task; its 67 critical-path
Reasoning calls are all escalations, while the Reasoning model planned in the background.

\subsection{AppMemory Under Routed Execution}
\begin{figure*}[t]
    \centering
    \includegraphics[width=\linewidth]{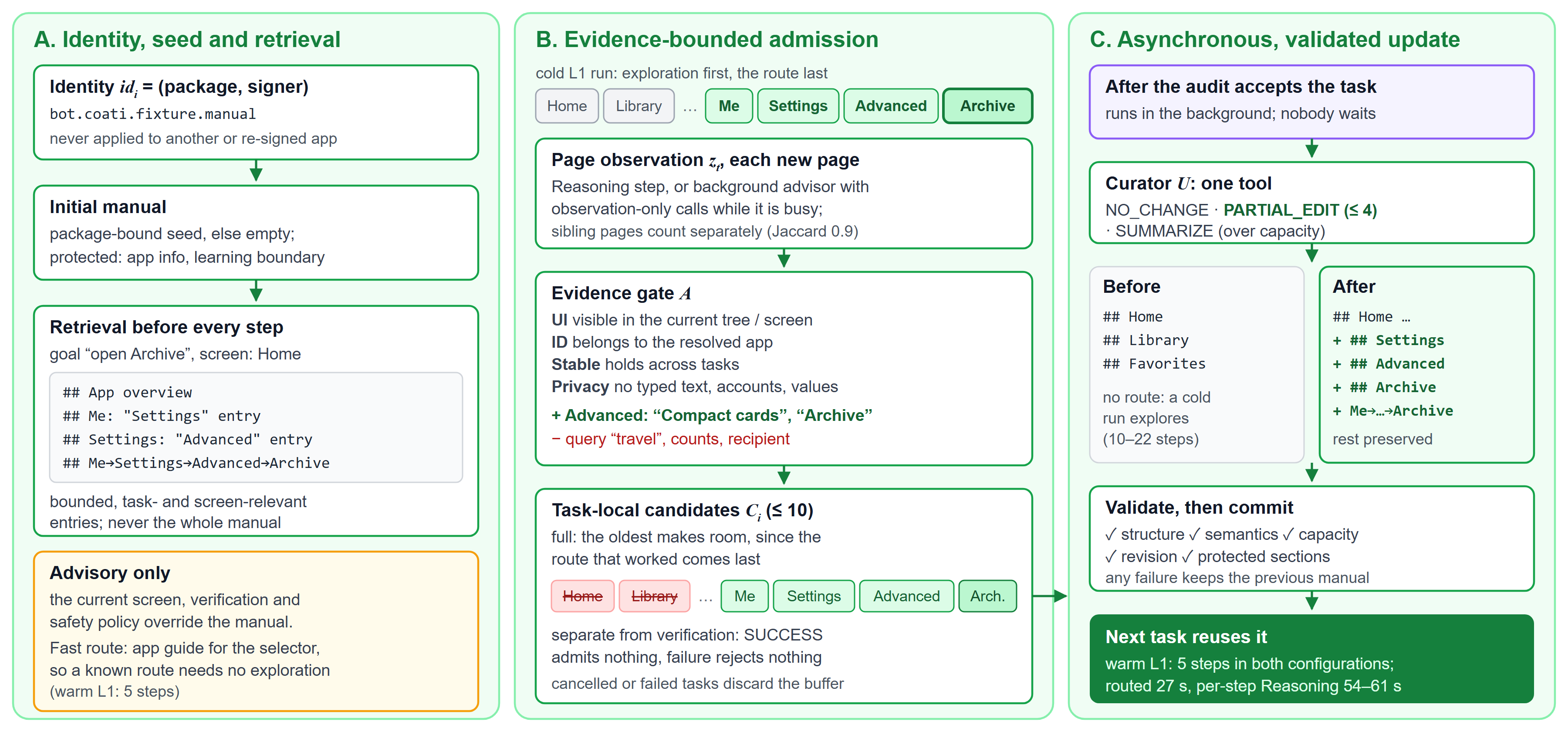}
    \caption{AppMemory on the learning benchmark (Manual Lab, task L1). (A) The manual is bound to
    the application identity and retrieved as bounded, advisory entries; on the fast route it is the
    selector's app guide. (B) Page observations pass the evidence gate, and the task-local buffer
    keeps the latest candidates, where the route that worked
    (Me~$\rightarrow$~Settings~$\rightarrow$~Advanced settings~$\rightarrow$~Archive) appears
    after exploration. (C) After the audit accepts the task, a curator applies a bounded edit that
    is committed only if validation succeeds; warm runs then take five steps.}
    \label{fig:appmemory}
\end{figure*}

The learning sequence of Section~\ref{subsec:rq4} on the phone runs L1 three times, then L3, with
routed execution and app knowledge in learning mode; L3 took 5 steps (23\,s).

\subsection{Case Study: A Commit Stopped by the Pre-commit Gate}
\label{subsec:case-precommit}

Task O1 asks to choose an item in a native order form, set a painted stepper to three copies, turn on
a painted toggle, fill two native text fields, and submit; hybrid perception merges a vision parse
for the painted widgets into the tree. In one routed phone run, the selector left the stepper at two
copies, completed the rest of the form and chose \emph{Submit order}. Before the tap, the pre-commit
gate asked the Fast model whether the screen satisfied the task and received ``Copies is set to 2
instead of 3''. The escalated Reasoning step tapped ``+'' once more, its verification confirmed three
copies, and the order was submitted with the requested values. Without the gate, only the completion
audit could have caught the wrong order, after the fact.

\end{document}